\documentclass[12pt]{article}
\def\Dirac{{\rm i}D\!\!\!\!/}

\def\e{{\rm e}}

\def\rmi{{\rm i}}
\def\rmd{{\rm d}}

\newcommand{\beq}{\begin{eqnarray}}
\newcommand{\eeq}{\end{eqnarray}}
\newcommand{\nn}{\nonumber}
\newcommand{\no}{\nonumber\\}
\begin{document}
%
%\begin{frontmatter}

\title{On the mistery of the missing pie in graphene}
\maketitle
\begin{center}
{Paola Giacconi}

%\address
{Istituto Nazionale di Fisica Nucleare, \\
Sezione di Bologna, 40126 Bologna, Italy}
%\footnote{Email address: paola.giacconi@bo.infn.it} 

\medskip
{Roberto Soldati}
%\footnote{Email address: roberto.soldati@bo.infn.it}

%\address
{Dipartimento di Fisica, Universit\`a di Bologna,\\
Istituto Nazionale di Fisica Nucleare, \\
Sezione di Bologna, 40126 Bologna, Italy}
%\address
%\date{\today}
%\maketitle
\end{center}
\begin{abstract}
%% Text of abstract
We investigate 
in some detail the structure of the electromagnetic current density
for the pseudo-relativistic massless spinor effective model for graphene.
It is shown that the pseudo-relativistic massless Dirac field theory 
in {\em 2+1} space-time dimensions and in the presence of a constant homogeneous electric
field actually leads to the measured current density and to the minimum 
quantum conductivity.
\end{abstract}
%
%\begin{keyword}
%\end{keyword}
%
%\end{frontmatter}
%\date{\today}
%
\section{Introduction}
One among the many far intringuing features of graphene is that \cite{geim}
its conductivity at zero magnetic field doesn't vanish close to the 
charge carriers neutrality point of concentration $n\le10^{-\,11}$ cm$^{-2}\,.$
Actually, on the one hand, the measured values for the lowest conductivity $\sigma_{\rm min}$
near the neutrality point are very close to the conductivity quantum
$e^2/h$ per carrier type \cite{nature1}. This zero-field quantum conductivity
does not apparently depend on the chemical potential and keeps
its single carrier value down to liquid helium temperatures \cite{geim}.
On the other hand, it turns out that
the insofar known theoretical predictions fail \cite{library}
in reproducing $\sigma_{\rm min}$, a disagreement known as {\em the mystery of the missing pie},
but for a very recent calculation \cite{bgss2}.
There, however, the deduction for the actual value of $\sigma_{\rm min}$
is certainly elegant and correct but
admittedly rather formal. In particular, the direction in the graphene's sample plane
of the minimum quantum current density is not at all clear from the above mentioned
theoretical derivation.

\medskip\noindent
As a further important ambipolar electric field effect,
which was not explained in \cite{bgss2},
it turns out that the single layer graphene
resistivity $\rho$ decreases rapidly to zero with adding charge carriers,
showing their very high-mobility \cite{geim}.
It is the aim of the present short note to fulfill this gap,
by developing a new derivation for the canonical quantum current density, 
within the  well
established effective model \cite{semenoff,mele} of a massless Dirac field theory in {\em 2+1} space-time dimensions
and in the presence of constant and homogeneous electromagnetic classical background
fields. In fact, the latter field theoretic model appears to be exactly solvable, 
in such a manner that
all the gauge invariant physical observables, like e.g. the mimimum quantum conductivity,
can be explicitly calculated and eventually compared with experimental data.

\medskip\noindent
In ref.~\cite{bgss2} it has been shown that the Euclidean effective action, or the
grand potential in statistical mechanics language, in the presence of a background uniform electrostatic field
turns out to be independent from the (complex) chemical potential and from the temperature.
Conversely, in the presence of a uniform magnetic field, the dependence of the effective action
upon the chemical potential is crucial in order to reproduce the measured values for the quantum Hall
conductivity~\cite{bgss1}. It follows thereby that in the present note we can restrict ourselves, without 
any loss of generality, to zero temperature and zero chemical potential planar quantum electrodynamics
in a constant homogeneous electric field. 
Hence, according to the leading order expansion around the 
Dirac points at the corner of the Brillouin zone in graphene
\cite{gusynin},
the classical pseudo-relativistic lagrangian in the Nikishov \cite{niki} 
temporal gauge $A^{\,\mu}=(0,-\,Ect,0)$ reads
\beq
{\mathcal L}={\bar\psi\,}(x)
\left(i\hbar\partial_{\,\mu}+\frac{e}{c}\,A_{\,\mu}\right)
\frac{\widetilde\gamma^{\,\mu}}{v_F}\;\psi(x)
\qquad\quad\bar\psi=\psi^{\,\dagger}\,\sigma_1
\eeq
Here $(-\,e)$ is the electron charge, $c$  the light velocity in vacuum,
$v_F\approx1/300$ is the Fermi velocity of the
massless Dirac quasiparticles in graphene, $\widetilde\gamma^{\,\mu}$
are the rescaled Dirac matrices, i.e.,
\[
\widetilde\gamma^{\,0} = c\,\gamma^{\,0}\qquad\quad
\widetilde\gamma^{\,\ell} = v_F\,\gamma^{\,\ell}\qquad\quad
(\,\ell=1,2\,)
\]
while $F_{01}=E_x=E>0$ denotes the constant and homogeneous
electric field pointing towards the positive $OX$ axis.
We remark that in the planar QED on the {\em 2+1} dimensional Minkowski space-time
the spinor field has canonical dimensions of the inverse of a length, i.e. of a 
wave number.
\section{Uniform electric field}%%%%%%%%%%%%%%%%%%%%%%%%%%%%%%%%%%%%%%%%%%%%%%%%%%%%%%%%%%%%%%%%%%%%%%%
Consider now a graphene sample in the presence of a constant electrostatic field
pointing towards the positive
$OX$ axis, i.e.,
\[
F_{01}=F^{\,10}=E_x=E\qquad\quad(\,E>0\,)
\]
In this section, we will choose a
gauge which leads to non-stationary sets of solutions.
In particular,
to solve the Dirac equation in the present {\em 2+1} dimensional massless case,
it is convenient to employ the representation  for the
gamma matrices
\beq
\gamma^0=\sigma_1\qquad\quad\gamma^1=i\sigma_2\qquad\quad\gamma^2=i\sigma_3
\eeq
After setting
$(x^0,x^1,x^2)=(ct,x,y)$
we get the massless Dirac operator
in the Nikishov temporal gauge
$A_{\,\mu}=(0,Ect,0)\,,$ i.e.,
\beq
\Dirac = \frac{i\hbar}{\upsilon_F}\,\partial_t\,\sigma_1+(i\hbar\,\partial_x+eEt)\,i\sigma_2 - \hbar\,\partial_y\,\sigma_3
\label{masslessdirac1}
\eeq
It follows that  in the Nikishov temporal gauge the 1-particle hamiltonian
\beq
H_t=\hbar\upsilon_F\left[\,\sigma_3(\rmi\,\partial_x-eEt/\hbar)-
\rmi\,\partial_y\,\sigma_2\,\right]
\eeq
is explicitly time dependent, which means that the energy does not lead to 
a good quantum number for the specification of the Fock space states of
the quantized massless Dirac spinor field.
Thus, at variance with the magnetic field case, in the present context
the Dirac hamiltonian
does not play here any key role.
Notice that
The classical Action for the massless Dirac field in the {\it 2+1}
dimensional Minkowski space-time and in the presence of a uniform background electric field takes
the form
\beq
{\mathcal A}= c\int_{-\infty}^\infty\rmd t\int\rmd{\bf r}\;
\bar\Psi(t,{\bf r})\left[\,
\frac{i\hbar}{\upsilon_F}\,\partial_t\,\sigma_1+(i\hbar\,\partial_x+eEt)\,i\sigma_2 - \hbar\,\partial_y\,\sigma_3\,\right]
\Psi(t,{\bf r})\nonumber
\eeq
whence it is apparent that the canonical physical dimensions of the spinor field are equal to the inverse of a length.
In order to obtain the solutions in this gauge,
it is convenient to introduce the spatial Fourier transforms
\beq
\Psi(t,x,y)={1\over2\pi}\int_{-\infty}^{\infty}{\rmd k}\int_{-\infty}^{\infty}\rmd p\
\e^{\,ipx+iky}\;\widetilde \Psi(t,p,k)
\label{fourier}
\eeq
with ${\bf p}=(p_x,p_y)\equiv(p,k)$ as well as the quantum electric length
\[
\ell_E\,\equiv\,\sqrt{\hbar\upsilon_F\over eE}
\]
and the dimensionless quantities
\beq
\xi &\equiv& \left(\,p-\frac{eEt}{\hbar}\,\right)\,\ell_E
\qquad\quad
\lambda\,\equiv\,k^2\,\ell_E^{\,2}
\eeq
so that the Dirac operator can be recast in the suitable form
\beq
\Dirac=\frac{\hbar}{\ell_E}\left\lgroup
\begin{array}{cc}
-\,ik\ell_E & -\,i\rmd_\xi-\xi\\
-\,i\rmd_\xi+\xi & \;ik\ell_E
\end{array}\right\rgroup
\eeq
If we set
\beq
\Psi(t,x,y)=\left\lgroup\begin{array}{c}
\varphi(t,x,y)\\
\chi(t,x,y)\end{array}\right\rgroup\qquad\quad
\widetilde \Psi(t,p,k)=\left\lgroup\begin{array}{c}
\widetilde\varphi(t,p,k)\\
\widetilde\chi(t,p,k)\end{array}\right\rgroup
\eeq
we obtain the coupled differential equations
\beq
\left\lbrace\begin{array}{c}
ik\ell_E\,\widetilde\varphi + (id_\xi+\xi)\,\widetilde\chi=0\\
(-id_\xi+\xi)\,\widetilde\varphi + ik\ell_E\,\widetilde\chi=0
\end{array}\right.
\eeq
Then, we can write
\beq
\left\lbrace\begin{array}{c}
\widetilde\chi=(\,i/k\ell_E)\,(-id_\xi+\xi)\,\widetilde\varphi\\
(d_\xi^{\,2}+\xi^2+\lambda+i)\,\widetilde\varphi=0
\end{array}\right.\qquad\quad\quad(\,k\not=0\,)
\label{normalmodes}
\eeq
\beq
\left\lbrace\begin{array}{c}
(id_\xi+\xi)\,\widetilde\chi=0\\
(-id_\xi+\xi)\,\widetilde\varphi=0
\end{array}\right.\qquad\quad\quad\qquad\quad\
(\,k=0\,)
\label{longitudinalmodes}
\eeq
Thus, in the presence of a homogeneous electric static field, one actually finds
the following
complete and orthonormal sets
\footnote{The complete and orthonormal sets of solutions listed below do coincide with
those ones obtained in \cite{bgss2} up to the overall dimensional factor $\ell_E^{\,-\,1}\,.$} 
of time dependent solutions of the Dirac equation : namely,
for $k\not=0$ we have either the {\sl outgoing normal modes}
\beq
u_{\,{\bf p}\,+}\,(t,{\bf r}) &=& {1\over2\pi\ell_E}\,\,
\exp\left\{i\,{\bf p}\cdot{\bf r} -\textstyle\frac18\,\pi\lambda\right\}\no
&\times& \left\lgroup\begin{array}{c}
{\textstyle\frac12}\,(1+i)\sqrt\lambda\,D_{\,i\lambda/2-1}(-\,z_{\,-})\\
D_{\,i\lambda/2}\,(-\,z_{\,-})
\\%\,[\,(1+i)\,\xi\,]
\end{array}\right\rgroup
\label{outparticle}
\eeq
which correspond to the normal modes of an outgoing quasiparticle
(massless electron or negatively charged neutrino) with momentum $\bf p$ and charge $-\,e\,,$
while
\beq
v_{\,{\bf p}\,+}\,(t,{\bf r}) &=& {1\over2\pi\ell_E}\,\,
\exp\left\{i\,{\bf p}\cdot{\bf r} -\textstyle\frac18\,\pi\lambda\right\}\no
&\times& \left\lgroup\begin{array}{c}
D_{-i\lambda/2}\,(-\,z_{\,+})\\
-\,{\textstyle\frac12}\,(1-i)\sqrt\lambda\,D_{-i\lambda/2-1}(-\,z_{\,+})\\
\end{array}\right\rgroup
\label{outantiparticle}
\eeq
do correspond to the normal modes of an outgoing antiquasiparticle (hole)
with momentum $\bf p$ and charge $e\,,$
where we have set
\[
z_\pm\,\equiv\,(1\pm i)\,\xi
\]
Furthermore the {\sl incoming normal modes} will be given by
\beq
u_{\,{\bf p}\,-}\,(t,{\bf r}) &=& {1\over2\pi\ell_E}\,\,
\exp\left\{i\,{\bf p}\cdot{\bf r} -\textstyle\frac18\,\pi\lambda\right\}\no
&\times& \left\lgroup\begin{array}{c}
-\,{\textstyle\frac12}\,(1+i)\,\sqrt\lambda\;D_{\,i\lambda/2-1}(z_{\,-})\\
D_{\,i\lambda/2}\,(z_{\,-})
\end{array}\right\rgroup
\label{inantiparticle}
\eeq
which correspond to the normal modes of an incoming quasiparticle
(massless electron or negatively charged neutrino) with momentum $\bf p$ and charge $-\,e\,,$
while
\beq
v_{\,{\bf p}\,-}\,(t,{\bf r}) &=& {1\over2\pi\ell_E}\,\,
\exp\left\{i\,{\bf p}\cdot{\bf r} -\textstyle\frac18\,\pi\lambda\right\}\no
&\times& \left\lgroup\begin{array}{c}
D_{-i\lambda/2}\,(z_{\,+})\\
{\textstyle\frac12}\,(1-i)\,\sqrt\lambda\;D_{-i\lambda/2-1}(z_{\,+})
\end{array}\right\rgroup
\label{inparticle}
\eeq
do correspond to the normal modes of an incoming antiquasiparticle (hole)
with momentum $\bf p$ and charge $e\,.$

Finally,
for $p_y=k=0$ we come to the so called {\sl longitudinal normal modes}
or {\sl zero modes}
\beq
u_{\,p\,\pm}\,(t,x) &=& {1\over\sqrt{2\pi}\,\ell_E}\,\,
\exp\left\{i\,{px} +\,{i\over2}\,\xi^2(t)\right\}
\left\lgroup\begin{array}{c}
0\\
1\end{array}\right\rgroup\;\equiv\;u_{\,p}\,(t,x) 
\label{longitudinalparticle}
\eeq
\beq
v_{\,p\,\pm}\,(t,x) &=& {1\over\sqrt{2\pi}\,\ell_E}\,\,
\exp\left\{i\,{px} -\,{i\over2}\,\xi^2(t)\right\}
\left\lgroup\begin{array}{c}
1\\
0\end{array}\right\rgroup\;\equiv\;v_{\,p}\,(t,x) 
\label{longitudinalantiparticle}
\eeq
The above sets of time dependent massless Dirac spinors have canonical physical dimensions
of a wave number, as it does, and are normalized in order to satisfy the following orthonormality
relations, viz.,
\beq
\int\rmd{\bf r}\;u_{\,{\bf q}\,\pm}^{\,\dagger}\,(t,{\bf r})\;u_{\,{\bf p}\,\pm}\,(t,{\bf r})=
\ell_E^{\,-2}\;\delta({\bf p}-{\bf q})
\eeq
\beq
\int\rmd{\bf r}\;v_{\,{\bf q}\,\pm}^{\,\dagger}\,(t,{\bf r})\;v_{\,{\bf p}\,\pm}\,(t,{\bf r})=
\ell_E^{\,-2}\;\delta({\bf p}-{\bf q})
\eeq
\beq
\int\rmd{\bf r}\;u_{\,{\bf q}\,\pm}^{\,\dagger}\,(t,{\bf r})\;v_{\,{\bf p}\,\pm}\,(t,{\bf r})=0
\qquad\quad\forall\,{\bf p},{\bf q}\in{\mathbf R}^2
\eeq
\beq
\int_{-\infty}^\infty\rmd x\;u_{\,q}^{\dagger}\,(t,x)\,u_{\,p}\,(t,x)=
\ell_E^{\,-2}\;\delta(p-q)
\eeq
\beq
\int_{-\infty}^\infty\rmd x\;v_{\,q}^{\dagger}\,(t,x)\,v_{\,p}\,(t,x)=
\ell_E^{\,-2}\;\delta(p-q)
\eeq
\beq
\int_{-\infty}^\infty\rmd x\;u_{\,q}^{\dagger}\,(t,x)\,v_{\,p}\,(t,x)=0
\qquad\quad\forall\,p,q\in{\mathbf R}
\eeq
\beq
\int\rmd{\bf r}\;u_{\,{\bf q}\,\pm}^{\,\dagger}\,(t,{\bf r})\;u_{\,p\,\pm}\,(t,x)=0
=\int\rmd{\bf r}\;u_{\,{\bf q}\,\pm}^{\,\dagger}\,(t,{\bf r})\;v_{\,p\,\pm}\,(t,x)
\eeq
\beq
\int\rmd{\bf r}\;v_{\,{\bf q}\,\pm}^{\,\dagger}\,(t,{\bf r})\;v_{\,p\,\pm}\,(t,x)=0
=\int\rmd{\bf r}\;v_{\,{\bf q}\,\pm}^{\,\dagger}\,(t,{\bf r})\;u_{\,p\,\pm}\,(t,x)
\eeq
The electric current density vector is provided by the invariance under U(1) phase transformations,
according to the Noether theorem, and reads
\beq
c\,J_0(t,{\bf r})=\varrho(t,{\bf r})={\rm q}\,\Psi^{\,\dagger}(t,x,y)\,\Psi(t,x,y)\\
J_x(t,{\bf r})={\rm q}\upsilon_F\,\bar\Psi(t,x,y)\,\gamma^1\,\Psi(t,x,y)
=-\,{\rm q}\upsilon_F\,\Psi^{\,\dagger}(t,x,y)\,\sigma_3\,\Psi(t,x,y)\\
J_y(t,{\bf r})={\rm q}\upsilon_F\,\bar\Psi(t,x,y)\,\gamma^2\,\Psi(t,x,y)
={\rm q}\upsilon_F\,\Psi^{\,\dagger}(t,x,y)\,\sigma_2\,\Psi(t,x,y)
\eeq
where q is the charge. For example, an outgoing particle
(massless electron or negatively charged neutrino)
of momentum ${\bf p}=(p,k)\,,\ k\not=0\,,$ and negative charge $-\,e$ carries the 
spatially homogeneous current density
\beq
\varrho&=& -\,e\;u_{\,{\bf p}\,+}^{\,\dagger}\,(t,{\bf r})\,u_{\,{\bf p}\,+}\,(t,{\bf r})
=\;{-\,e^2E\over2\pi h\upsilon_F}\\
J_x(t\,;{\bf p})&=&-\,{e^2E\over2\pi h}\,\exp\left\{-\textstyle\frac14\,\pi\lambda\right\}\no
&\times&\left(\,|\,D_{\,i\lambda/2}\,(-\,z_{\,-})\,|^{\,2}-
{\textstyle\frac12}\,\lambda\,|\,D_{\,i\lambda/2-1}(-\,z_{\,-})\,|^{\,2}\,\right)\\
J_y(t\,;{\bf p})&=& {ie^2E\over2\pi h}\,\exp\left\{-\textstyle\frac14\,\pi\lambda\right\}\no
&\times&D_{\,i\lambda/2}\,(-\,z_{\,-})\,\,{\textstyle\frac12}\,(1-i)\sqrt\lambda\,D_{-i\lambda/2-1}(-\,z_{\,+})\
+\ {\rm c.c.}
\eeq
Now, since the transverse momentum of a normal mode solution is different from zero by definition, 
it turns out that in the weak field limit $\lambda\to\infty$ the current density components of
all the normal modes become
exponentially small, for any outgoing and/or  incoming particle and/or antiparticle
normal mode massless Dirac spinor. Conversely, for the remaining set of longitudinal normal modes
with $p_y=k=0$ we immediately find for quasiparticles
\beq
\varrho
=-\;{e^2E\over h\upsilon_F}\,,\qquad\quad
J_x = -\;{e^2E\over h}\,,\qquad\quad J_y=0\,,
\qquad\quad\forall\,p\in{\mathbf R}
\eeq
while for antiquasiparticles we obviously find
\beq
\varrho
= {e^2E\over h\upsilon_F}\,,\qquad\quad
J_x = {e^2E\over h}\,,\qquad\quad J_y=0\,,
\qquad\quad\forall\,p\in{\mathbf R}
\eeq
This actually means that the only observable nonvanishing current density, in the weak field limit,
is the longitudinal one,
so that the minimum quantum conductivity\footnote{Notice that in physical units we have
${e^2}/{h}\simeq3.5\times10^7\ {\rm cm\ s}^{-1}\;\approx\;4\times10^{-5}\ \Omega^{-1}$}
for a graphene sample in the presence of a uniform electric field is provided by
$\sigma_{\,\rm min}=4e^2/h\simeq\;1.4\times10^8\ {\rm cm\ s}^{-1}\;\simeq\;1.5\times10^{-4}\ \Omega^{-1}$ as 
indeed observed experimentally, which entails
$\rho_{\,\rm max}\;\simeq\;6.6\ {\rm K}\Omega\,,$ where the factor four
is due \cite{gusynin,castro} to the two inequivalent irreducible two-dimensional spinor representations,
as well as to the so called {\sl valley degeneracy}.
We remark that the present value of 
$\sigma_{\,\rm min}$ is precisely the very same obtained in \cite{bgss2},
but from a quite different method based upon the Euclidean effective action.

On the other side, i.e. in the strong field limit $\lambda\to 0\,,$ we readily get
for graphene quasiparticles
\beq
\varrho(\lambda=0)
=\left\lbrace
\begin{array}{cc}
{-\,2e^2E /\pi h\upsilon_F}\;, &\forall\,{\bf p}\in{\mathbf R}^2\vee k\not=0\\
{-\,4e^2E / h\upsilon_F}\;, &\forall\,{p}\in{\mathbf R}\vee k=0
\end{array}\right.\\
J_x(\lambda=0) =\left\lbrace
\begin{array}{cc}
 -\;{2e^2E /\pi h}\;, &\forall\,{\bf p}\in{\mathbf R}^2\vee k\not=0\\
-\;{4e^2E / h}\;, &\forall\,{p}\in{\mathbf R}\vee k=0
\end{array}\right.\\
J_y(\lambda=0)=0\;,\qquad\quad\forall\,{\bf p}\in{\mathbf R}^2
\eeq
while for antiquasiparticles we obviously find the opposite signs.
This means that {\bf all the quantum states} give 
leading constant contributions, in the strong field regime,
to the conductivity of a graphene single layer sample, i.e.,
$\mp\,4e^2/2\pi h$ for the quasiparticle/antiquasiparticle normal modes with $k\not=0\,,$ 
while $\mp\,4e^2/ h$ for the corresponding zero modes.
This feature nicely explain the rapid fall down of graphene resistivity
with increasing the charge carriers density, viz., their very high mobility.
As a matter of fact, if there are $N$ electrons in the graphene sample,
for instance, then the leading resistivity for a sufficiently strong
electric field becomes
\beq
\rho\,\buildrel \lambda\,\to\,0\over\sim\,\frac{-\,h}{4Ne^2}\,\times\,\left\lbrace
\begin{array}{cc}
1  & {\rm for}\ k=0\\
2\pi & {\rm for}\ k\not=0
\end{array}\right.
\eeq
which is vanishing for e.g. a typical density $n=10^{19}$ electrons cm$^{-2}\,,$
as experimentally observed.
\section{Discussion and conclusions}
In this note we have analysed the exact current density
of graphene within the pseudo-relativistic effective field theoretic model
of a massless Dirac field in
a {\em 2+1} dimensional space-time under the influence of a uniform background
electromagnetic field. We have shown that, in the limit of a vanishing electric field,
a minimum quantum conductivity $\sigma_{\,\rm min}$ does survive, which is entirely due to the longitudinal
zero modes, i.e. with $k=0$, and the actual value of which is in agreement with the
experimental finding. Moreover, it has been proved that in the strong field regime
all the charged graphene quantum states provide a leading, dominant, constant contribution to 
the conductivity, a feature that simply explains the high charge carriers mobility and the rapid fall down of
resistivity with increasing number of the charge carriers.
This is the main prediction of the massless, pseudo-relativistic planar QED
effective model for graphene, in the presence of a constant homogeneous electic field.
In turn, the related minimum quantum conductivity perfectly reproduces 
its measured values, at variance with many previously obtained different results \cite{library}
based upon the Kubo formula.
The ultimate reason for this might be that our exact solutions are truly non-perturbative, viz., they do not
reproduce the free field spinor when $E\to0\,,$ whilst the Kubo approach is a 
perturbative linear response approximation.
\section*{Acknowledgements}
We are grateful to N. Protasov 
for a fruitful and stimulating correspondence. We warmly thank Carlota Gabriela Beneventano and
Eve Mariel Santangelo for a careful reading of the manuscript and enlightening comments.
We wish to acknowledge the support of the Istituto Nazionale di Fisica Nucleare,
Iniziativa Specifica PI13, that contributed to the successful completion of this project.
\end{document}